\documentclass[showpacs,aps,graphicx,twocolumn]{revtex4}
\usepackage{graphicx}

\begin{document}

\title{ Dynamics of entanglement for a two-parameter class of states
in a qubit-qutrit system\footnote{Published in Commun. Theor. Phys.
\textbf{57},  983-990 (2012)}}
\author{Hai-Rui Wei, Bao-Cang Ren, Tao Li, Ming Hua, and Fu-Guo Deng\footnote{{Corresponding author. Email address:
fgdeng@bnu.edu.cn}}}
\address{  Department of Physics, Applied Optics Beijing Area Major Laboratory, Beijing Normal University, Beijing 100875, China }
\date{\today }

\begin{abstract}
We investigate the dynamics of entanglement for a two-parameter
class of states in a hybrid qubit-qutrit system under the influence
of various dissipative channels. Our results show that entanglement
sudden death (ESD) is a general phenomenon and it usually takes
place in a qubit-qutrit system interacting with various noisy
channels, not only the case with dephasing and depolarizing channels
observed by others. ESD can only be avoided for some initially
entangled states under some particular noisy channels. Moreover, the
environment affects the entanglement and the coherence of the system
in very different ways.
\end{abstract}
 \maketitle

\section{Introduction}
\label{sec1}

Quantum entanglement is  a vital resource for quantum information
processing \cite{r1}. However, isolating a quantum system completely
from its environment is plainly an impossible task and each quantum
system will inevitably interact with its environment. Therefore, it
is important to investigate the behavior of an entangled quantum
system under the influence of its environment. Recently, Yu and
Eberly \cite{r7,r8}  investigated the dynamics of two-qubit
entangled states undergoing various modes of decoherence. They found
that it takes an infinite time to complete decoherence locally, the
global entanglement may be lost in  a finite time, and the decay of
a single-qubit coherence can be slower than the decay of a two-qubit
entanglement. The abrupt disappearance of entanglement in a finite
time was named ''entanglement sudden death''(ESD). A geometric
interpretation of the phenomenon is given in Ref.\cite{Geometric}.
In addition, experimental evidences of ESD have been reported for
optical setups \cite{Optical1,Optical2}  and atomic ensembles
\cite{Atomic}.  Clearly, ESD can seriously affect the applications
of entangled states in  a practical quantum information processing.
Recently,  dynamics of entanglement has received increasing
attention \cite{nomar,zhaopra1,zhaopra2}.

ESD in  finite-dimensional systems is not limited only to two-qubit
systems. It may be occurs in a composite quantum system with a
larger dimension  and a multiqubit system  as well
\cite{Large1,Large2,r9,r10,mul1,mul2,mul3}.  The dissipative
dynamics for a specific one-parameter class of states in a
qubit-qutrit ($2\otimes3$) system interacting with dephasing and
depolarizing channels was studied by Ann \emph{et al.} \cite{r9} and
Khan \cite{r10}, respectively. Ann \emph{et al.} \cite{r9}
conjectured that ESD exists in all bipartite quantum systems. Khan
\cite{r10} showed that no ESD happens in any density matrix of a
qubit-qutrit system when only the qubit is coupled to its local
depolarizing channel but the re-birth of entanglement occurs in
particular initial states. However, for general qubit-qutrit states
and other common noise channels, the dissipative dynamics of a
hybrid qubit-qutrit is not presented.

In this paper, we devote to investigate the behavior of entanglement
for a two-parameter class of states in a qubit-qutrit system under
the influence of both two independent (multi-local) and only one
(local) various noise channels, such as dephasing, phase-flip,
bit-(trit-) flip, bit-(trit-) phase-flip, and depolarizing channels.
Using negativity for quantifying entanglement, some analytical or
numerical results are presented. We find that ESD is a general
phenomenon in a  qubit-qutrit system undergoing all these noise
channels, not only the case with dephasing and depolarizing channels
observed by others \cite{r9}.  It is interesting to show that ESD
always takes place in any density matrix when each subsystem couples
to its depolarizing channel or only the qutrit couples to its
trit-flip or trit-phase-flip channels. ESD can only be avoided in
some initial states undergoing particular noise channels. For
example, no ESD occurs when the system under the influence of
multi-local (local) dephasing, multi-local (local) phase-flip, local
bit-flip, and local bit-phase-flip channels if it is initially in
the state shown in Eq.(\ref{eq.1}) with the parameter $b=0$.  Our
results show that the noise channels affect the entanglement and the
coherence of a hybrid qubit-qutrit system in very different ways.
For local or multi-local dephasing, phase-flip, and depolarizing
noise channels, a time scale of disentanglement is usually shorter
than the decay of the off-diagonal dynamics, and coherence
disappears in an infinite-time limit. For multi-local and local
bit-flip and bit-phase-flip channels, disentanglement  occurs in an
infinite time, but coherence does not disappear even though
$t\mapsto\propto$.

This paper is organized as follows. In Sec.\ref{sec2}, we motivate
the choice of a two-parameter class of states in a qubit-qutrit
system, and the physical model  are introduced. In Sec.\ref{sec3},
entanglement dynamics of a two-parameter class of states in a
qubit-qutrit system under the influence of  local and multi-local
dephasing, phase-flip, bit-(trit-) flip, bit-(trit-) phase-flip, and
depolarizing noise channels are discussed, respectively. Discussion
and summary are shown in Sec.\ref{sec4}.

\section{ Initial states and noise model }\label{sec2} 

A two-parameter  class of states with real parameters in a hybrid
qubit-qutrit ($2 \otimes 3$) quantum system \cite{r15} can be
described as
\begin{eqnarray}
\rho_{bc}(0)&=&
a\left(|02\rangle\langle02|+|12\rangle\langle12|\right)+
b(|\phi^{+}\rangle\langle\phi^{+}|+ \nonumber
\\&&|\phi^{-}\rangle\langle\phi^{-}|+|\psi^{+}\rangle\langle\psi^{+}|)+c|\psi^{-}\rangle\langle\psi^{-}|, \label{eq.1}
\end{eqnarray}
where
\begin{eqnarray}
|\phi^{\pm}\rangle &=& \frac{1}{\sqrt{2}}(|00\rangle\pm|11\rangle),\nonumber\\
|\psi^{\pm}\rangle &=& \frac{1}{\sqrt{2}}(|01\rangle\pm|10\rangle),
\label{eq.2}
\end{eqnarray}
and  $a$, $b$, and $c$ are three real parameters, and they satisfy
the relation $2a+3b+c=1$. $\vert 0\rangle$ and $\vert 1\rangle$ are
the two eigenstates of a two-level quantum system (qubit) or the
eigenstates of a three-level quantum system (qutrit) with the other
eigenstate $\vert 2\rangle$. The two-parameter class of states
$\rho_{bc}(0)$ can be obtained from an arbitrary state of a $2
\otimes 3$ quantum system by means of local quantum operations and
classical communication \cite{r15}.

For an arbitrary mixed state $\rho^{AB}$ in a $2\otimes2$ or a
$2\otimes3$ system, its entanglement can well be characterized and
quantified by its
negativity,\textsuperscript{\cite{negativity1,negativity2}}
\begin{eqnarray}                                               \label{eq.3}
N(\rho^{AB})=\parallel\rho^{T_{B}}\parallel_{1}-1,
\end{eqnarray}
which corresponds to the absolute value of the sum of negative
eigenvalues of $\rho^{T_{B}}$(the partial transpose $\rho^{T_{B}}$
associated with an arbitrary product orthonormal basis $f_i \otimes
f_j$ is defined by the matrix elements:
$\rho_{m\mu,n\nu}^{T_B}\equiv \langle f_m \otimes
f_\mu|\rho^{T_B}|f_n \otimes f_\nu\rangle=\rho_{m\nu,n\mu}$), i.e.,
\begin{eqnarray}                                        \label{eq.4}
N(\rho^{AB})=2\max\{0,-\lambda_{S}\},
\end{eqnarray}
where $\lambda_{S}$ represents the sum of all negative eigenvalues
of $\rho^{T_{B}}$. $N(\rho^{AB})=0$ for an unentangled state.
Therefore, from Eq.(\ref{eq.1}) one can obtain the range of
parameters as $3b<c\leq1-3b$, i.e., $b\in[0,1/6)$  for the initial
entangled states.

In our physical model of noise for a qubit-qutrit system (composed
of a two-level subsystem $A$ and a three-level subsystem $B$), the
two subsystems interact with their  environments independently. The
evolved states of the initial density matrix of such a system when
it is influenced by multi-local environments can be given compactly
by
\begin{eqnarray}                                        \label{eq.5}
\rho_{bc}^{AB}(t)=\sum_{i=1}^{2}\sum_{j=1}^{3}F_{j}^{B}E_{i}^{A}\rho^{AB}_{bc}(0)E_{i}^{A\dagger}F_{j}^{B
\dagger}.
\end{eqnarray}
Here, the operators $E_{i}^{A}$ and $F_{j}^{B}$ are the Kraus
operators which are used to describe the noise channels acting on
the qubit $A$ and the qutrit $B$, respectively. They satisfy the
completeness relations $E_{i}^{A\dagger}E_{i}^{A}=I$ and $F_{j}^{B
\dagger}F_{j}^{B}=I$ for all $t$.

\bigskip
\section{Dynamics of entanglement under decoherence} \label{sec3}

It is important to consider the possible degradation of any
initially prepared entanglement due to decoherence. In this section
we investigate what happens to the entanglement in a qubit-qutrit
system under common noise channels for qubit (qutrit): dephasing,
phase-flip, bit-(trit-) flip, bit-(trit-) phase-flip, and
depolarizing channels. The two specific environment noise situations
will be considered: (i) local and (ii) multi-local. In the case (i),
only one part of a qubit-qutrit system ($S$) interacts with its
environment. In the case (ii), both the two parts of $S$ interact
with their local environments, independently.

\subsection{Dephasing channels} \label{sec3.1}

The set of Kraus operators for a single qubit $A$ and a single
qutrit $B$ that reproduce the effect of a dephasing channel are
given  by
\begin{eqnarray}                                                   \label{eq.6}
E_{1}^{A}&=&\left(\begin{array}{cc}
1&0\\
0&\sqrt{1-\gamma_{A}}\\
\end{array}
\right)\otimes I_{3}, \nonumber \\
E_{2}^{A}&=&\left(\begin{array}{cc}
0&0\\
0&\sqrt{\gamma_{A}}\\
\end{array}
\right)\otimes I_{3},\\                       \label{eq.7}
F_{1}^{B}&=&I_{2}\otimes\left(\begin{array}{ccc}
1&0&0\\
0&\sqrt{1-\gamma_{B}}&0\\
0&0&\sqrt{1-\gamma_{B}}\\
\end{array}
\right), \nonumber \\
F_{2}^{B}&=&I_{2}\otimes\left(\begin{array}{ccc}
0&0&0\\
0&\sqrt{\gamma_{B}}&0\\
0&0&0\\
\end{array}
\right), \nonumber \\
F_{3}^{B}&=&I_{2}\otimes\left(\begin{array}{ccc}
0&0&0\\
0&0&0\\
0&0&\sqrt{\gamma_{B}}\\
\end{array}
\right).
\end{eqnarray}
The time-dependent parameters are defined as
$\gamma_{A}=1-e^{-t\Gamma_{A}}$ and $\gamma_{B}=1-e^{-t\Gamma_{B}}$.
Here $\gamma_{A}$, $\gamma_{B}\in[0,1]$. $\Gamma_{A}$ ($\Gamma_{B}$)
denotes the decay rate of the subsystem $A$ ($B$).

According to Eq.(\ref{eq.5}), the time-dependent evolved density
operator $\rho^{AB}(t)$ of a hybrid qubit-qutrit system, which is
initially in the entangled state $\rho^{AB}(0)$, is given by
\begin{widetext}
\begin{center}
\begin{eqnarray}                                                   \label{eq.8}
\rho^{AB}(t)=\left(\begin{array}{cccccc}
b&0&0&0&0&0\\
0&\frac{b+c}{2}&0&\frac{(b-c)\sqrt{(1-\gamma_{A})(1-\gamma_{B})}}{2}&0&0\\
0&0&\frac{1-c-3b}{2}&0&0&0\\
0&\frac{(b-c)\sqrt{(1-\gamma_{A})(1-\gamma_{B})}}{2}&0&\frac{b+c}{2}&0&0\\
0&0&0&0&b&0\\
0&0&0&0&0&\frac{1-c-3b}{2}\\
\end{array}
\right).
\end{eqnarray}
\end{center}
\end{widetext}
In order to characterize the dynamics of evolution for  the density
matrix and consider the entanglement of this system quantified by
negativity, we should calculate the eigenvalues of the partial
transpose of the time-evolved density matrix $\rho^{AB}(t)$ and
determine the potential negative eigenvalues.

\textsl{(1) Multi-local dephasing channel}. The eigenvalue which can
potentially be negative is
\begin{eqnarray}                                                   \label{eq.9}
\lambda=b-\frac{c-b}{2}\sqrt{(1-\gamma_{A})(1-\gamma_{B})}.
\end{eqnarray}
The entanglement of the qubit-qutrit system under a multi-local
dephasing channel  is
\begin{eqnarray}                                                   \label{eq.10}
N^{mul-loc}(\rho^{AB})=2\max\{0,
\frac{c-b}{2}\sqrt{(1-\gamma_{A})(1-\gamma_{B})}-b\}.\nonumber\\
\end{eqnarray}
It is easy to obtain that all the states which are initially
entangled ($3b<c\leq1-3b$) become separable when
$(1-\gamma_{A})(1-\gamma_{B})\leq\left(\frac{2b}{c-b}\right)^2$ and
$b\neq0$.

\textsl{(2) Qubit dephasing channel only}. If the qutrit field is
turned off (i.e., $\gamma_{B}=0$), that is, only a dephasing noise
acts on qubit $A$ alone, the  entanglement of the qubit-qutrit
system is
\begin{eqnarray}                                                     \label{eq.11}
N^{bit}(\rho^{AB})=2\max\{0, \frac{c-b}{2}\sqrt{1-\gamma_{A}}-b\}.
\end{eqnarray}
All the states which are initially entangled ($3b<c\leq1-3b$) become
separable as soon as $\gamma_{A}\geq
1-\left(\frac{2b}{c-b}\right)^2$ and $b\neq0$.

\textsl{(3) Qutrit dephasing channel only}. By the same argument as
that made in the qubit dephasing noise, for a dephasing noise acting
on qutrit $B$ alone, the  entanglement is
\begin{eqnarray}                                                   \label{eq.12}
N^{trit}(\rho^{AB})=2\max\{0, \frac{c-b}{2}\sqrt{1-\gamma_{B}}-b\}.
\end{eqnarray}
A single-qutrit dephasing channel will induce ESD in the
qubit-qutrit system when $\gamma_{B}\geq
1-\left(\frac{2b}{c-b}\right)^2$ and $b\neq 0$.

Together with these pieces of results, one can see that ESD takes
place under local and multi-local dephasing channels in a general
qubit-qutrit system if and only if the system is initially in the
state shown in Eq.(\ref{eq.1}) with the parameter $b\neq0$.
Moreover, the smaller $c/b$, the longer the death time range.
However, if $b=0$,
no ESD occurs.\\

\subsection{ Phase-flip channels}\label{sec3.2} 

The Kraus operators describing the phase-flip channel for a single
qubit $A$ are given by
\begin{eqnarray}                                                   \label{eq.13}
E_{1}^{A}&=&\sqrt{1-\frac{\gamma_{A}}{2}}\left(\begin{array}{cc}
1&0\\
0&1\\
\end{array}
\right)\otimes I_{3}, \nonumber \\
E_{2}^{A}&=&\sqrt{\frac{\gamma_{A}}{2}}\left(\begin{array}{cc}
1&0\\
0&-1\\
\end{array}
\right)\otimes I_{3},
\end{eqnarray}
and those for a single qutrit $B$ can be written as
\begin{eqnarray}                                                   \label{eq.14}
F_{1}^{B}&=&I_{2}\otimes\sqrt{1-\frac{2\gamma_{B}}{3}}\left(\begin{array}{ccc}
1&0&0\\
0&1&0\\
0&0&1\\
\end{array}
\right), \nonumber \\
F_{2}^{B}&=&I_{2}\otimes\sqrt{\frac{\gamma_{B}}{3}}\left(\begin{array}{ccc}
1&0&0\\
0&e^{-i2\pi/3}&0\\
0&0&e^{i2\pi/3}\\
\end{array}
\right),  \nonumber \\
F_{3}^{B}&=&I_{2}\otimes\sqrt{\frac{\gamma_{B}}{3}}\left(\begin{array}{ccc}
1&0&0\\
0&e^{i2\pi/3}&0\\
0&0&e^{-i2\pi/3}\\
\end{array}
\right),
\end{eqnarray}
where $\gamma_{A}=1-e^{-t\Gamma_{A}}$,
$\gamma_{B}=1-e^{-t\Gamma_{B}}$, and
$\gamma_{A},\gamma_{B}\in[0,1]$. $\Gamma_{A}$ ($\Gamma_{B}$)
represents the decay rate of the subsystem $A$ ($B$).

We can obtain the time-evolved density-matrix dynamics
$\rho^{AB}(t)$ of the qubit-qutrit system under a phase-flip
channel, according to Eq.(\ref{eq.5}). That is,
\begin{widetext}
\begin{center}
\begin{eqnarray}                                                   \label{eq.15}
\rho^{AB}(t)=\left(\begin{array}{cccccc}
b&0&0&0&0&0\\
0&\frac{b+c}{2}&0&\frac{(b-c)(1-\gamma_{A})(1-\gamma_{B})}{2}&0&0\\
0&0&\frac{1-c-3b}{2}&0&0&0\\
0&\frac{(b-c)(1-\gamma_{A})(1-\gamma_{B})}{2}&0&\frac{b+c}{2}&0&0\\
0&0&0&0&b&0\\
0&0&0&0&0&\frac{1-c-3b}{2}\\
\end{array}
\right).
\end{eqnarray}
\end{center}
\end{widetext}
With the same argument as that made in dephasing channels, some
results can be presented as follows.


\textsl{(1) Multi-local phase-flip channel}. The entanglement of the
qubit-qutrit system under a multi-local phase-flip channel is
calculated as
\begin{widetext}
\begin{center}
\begin{eqnarray}                                                   \label{eq.16}
 &&N^{mul-loc}(\rho^{AB})= 2\max\left\{0,
\frac{(c-3b)+(c-b)((1-\gamma_{A})(1-\gamma_{B})-1)}{2}\right\}.\nonumber
\end{eqnarray}
\end{center}
\end{widetext}
That is,  all the states which are initially entangled
($3b<c\leq1-3b$) become separable for all values of
$(1-\gamma_{A})(1-\gamma_{B})\leq \frac{2b}{c-b}$ and $b\neq0$.

\textsl{(2) Qubit phase-flip channel only}. The entanglement of the
system is
\begin{eqnarray}                                                   \label{eq.17}
N^{bit}(\rho^{AB})=2\max\{0,\frac{c-3b-\gamma_{A}(c-b)}{2}\}.
\end{eqnarray}
All the states which are initially entangled ($3b<c\leq1-3b$) become
separable as soon as $\gamma_{A}\geq \frac{c-3b}{c-b}$  and $b \neq
0$.

\textsl{(3) Qutrit phase-flip channel only}. The negativity for all
the states which are initially entangled ($3b<c\leq1-3b$) is given
as
\begin{eqnarray}                                                   \label{eq.18}
N^{trit}(\rho^{AB})=2\max\{0, \frac{c-3b-\gamma_{B}(c-b)}{2}\}.
\end{eqnarray}
All the states which are initially entangled become separable when
$\gamma_{B}\geq \frac{c-3b}{c-b}$ and $b\neq 0$.

Similar to the case with a dephasing noise, ESD  also takes place
when the system undergos a phase-flip noise environment but no ESD
occurs when the parameter $b=0$. Difference from the dephasing
channel, the decay time range  under the multi-local phase-flip
channel is shorter than that in the local channel if $3b<c<5b$, and
longer for $5b<c\leq 1-3b$.

\bigskip

\subsection{ Bit-(Trit-) flip channels} \label{sec3.3}

The Kraus operators describing the bit-flip channel for a single
qubit $A$ are given by
\begin{eqnarray}                                                   \label{eq.19}
E_{1}^{A}&=&\sqrt{1-\frac{\gamma_{A}}{2}}\left(\begin{array}{cc}
1&0\\
0&1\\
\end{array}
\right)\otimes I_{3}, \nonumber \\
E_{2}^{A}&=&\sqrt{\frac{\gamma_{A}}{2}}\left(\begin{array}{cc}
0&1\\
1&0\\
\end{array}
\right)\otimes I_{3},
\end{eqnarray}
and those for a single qutrit $B$ can be written as
\begin{eqnarray}                                                   \label{eq.20}
F_{1}^{B}&=&I_{2}\otimes\sqrt{1-\frac{2\gamma_{B}}{3}}\left(\begin{array}{ccc}
1&0&0\\
0&1&0\\
0&0&1\\
\end{array}
\right), \nonumber \\
F_{2}^{B}&=&I_{2}\otimes\sqrt{\frac{\gamma_{B}}{3}}\left(\begin{array}{ccc}
0&0&1\\
1&0&0\\
0&1&0\\
\end{array}
\right), \nonumber \\
F_{3}^{B}&=&I_{2}\otimes\sqrt{\frac{\gamma_{B}}{3}}\left(\begin{array}{ccc}
0&1&0\\
0&0&1\\
1&0&0\\
\end{array}
\right),
\end{eqnarray}
where $\gamma_{A}=1-e^{-t\Gamma_{A}}$,
$\gamma_{B}=1-e^{-t\Gamma_{B}}$, and
$\gamma_{A},\gamma_{B}\in[0,1]$.

The matrix elements of $\rho$ after the interaction time $t$ under a
multi-local bit-(trit-) flip channel are given by

\begin{widetext}
\begin{center}
\begin{eqnarray}                                                   \label{eq.21}
\rho_{11}(t)&=\rho_{55}(t)=&\frac{1}{12}(12b+3(b-c)\gamma_{A}(\gamma_{B}-1)\nonumber\\&&+2(1-6b)\gamma_{B}),\nonumber\\
\rho_{22}(t)&=\rho_{44}(t)=&\frac{1}{12}(6(b+c)-3(b-c)\gamma_{A}(\gamma_{B}-1)\nonumber\\&&+(2-6b-6c)\gamma_{B}),\nonumber\\
\rho_{33}(t)&=\rho_{66}(t)=&\frac{1}{6}\left(3(1-3b-c)+(9b+3c-2)\gamma_{B}\right),\nonumber\\
\rho_{15}(t)&=\rho_{51}(t)=&\frac{1}{12}(b-c)\gamma_{A}(3-2\gamma_{B}),\nonumber\\
\rho_{35}(t)&=\rho_{53}(t)=&\rho_{16}(t)=\rho_{61}(t)=\frac{1}{12}(b-c)(2-\gamma_{A})\gamma_{B},\nonumber\\
\rho_{26}(t)&=\rho_{62}(t)=&\rho_{34}(t)=\rho_{43}(t)=\frac{1}{12}(b-c)\gamma_{A}\gamma_{B},\nonumber\\
\rho_{24}(t)&=\rho_{42}(t)=&\frac{1}{12}(b-c)(2-\gamma_{A})(3-2\gamma_{B}),
\end{eqnarray}
\end{center}
\end{widetext}
and all the remaining matrix elements are zero. Here
$\rho=(\rho_{ij})$  and $i,j=1,\ldots,6$.

\textsl{(1) Qubit bit-flip channel only}. We obtain the same results
as the case with a phase-flip channel.

\textsl{(2) Qutrit trit-flip channel only}. The negativity for all
the states which are initially entangled ($3b<c\leq1-3b$) can be
given by
\begin{widetext}
\begin{center}
\begin{eqnarray}                                                        \label{eq.22}
N^{trit}(\rho^{AB})=2\max\left\{0,\frac{3b-9c-(1-8b+2c)\gamma_{B}}{6}\right\}.
\end{eqnarray}
\end{center}
\end{widetext}
The states  become separable once $\gamma_{B}\geq
\frac{3c-9b}{1-8b+2c}$.

\textsl{(3) Multi-local bit-(trit-) flip channel}. For simplicity,
we choose the local asympotic bit-(trit-) flip rates
$\Gamma_{A}=\Gamma_{B}=\Gamma$ to discuss the effect of this noise
on entanglement. Unfortunately, it is difficult to calculate the
analytical eigenvalues of the partial transpose of the time-evolved
density matrix. Therefore, we take the numerical calculation for
various initial states as examples to investigate the dynamics of
entanglement of the qubit-qutrit system under a multi-local
bit-(trit-) flip channel.

\begin{figure}[!h]
\begin{center}
\includegraphics[width=8 cm,angle=0]{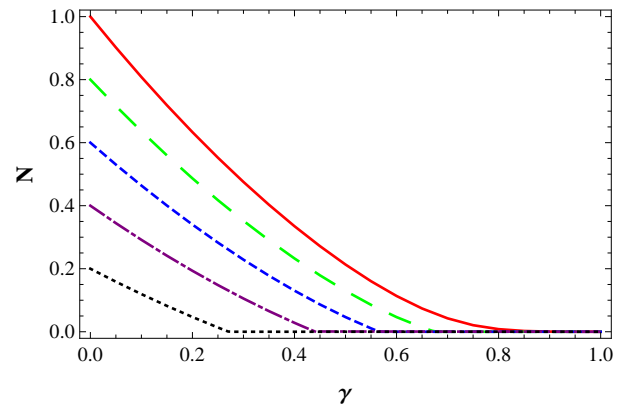}
\caption{Dynamics of entanglement for the system undergoing the
multi-local bit- (trit-) flip noise with $a=0$. $N$ represents the
negativity of a hybrid qubit-qutrit system. $\gamma$ is the
time-dependent parameter and
$\gamma_A=\gamma_B=\gamma=1-e^{-t\Gamma}$. The solid, dashed,
short-dashed, dashed-dotted, and dotted lines correspond to $b=0$,
$b=1/30$, $b=2/30$, $b=3/30$, and $b=4/30$, respectively.}
\label{Fig1}
\end{center}
\end{figure}

\begin{figure}[!h]
\begin{center}
\includegraphics[width=8 cm,angle=0]{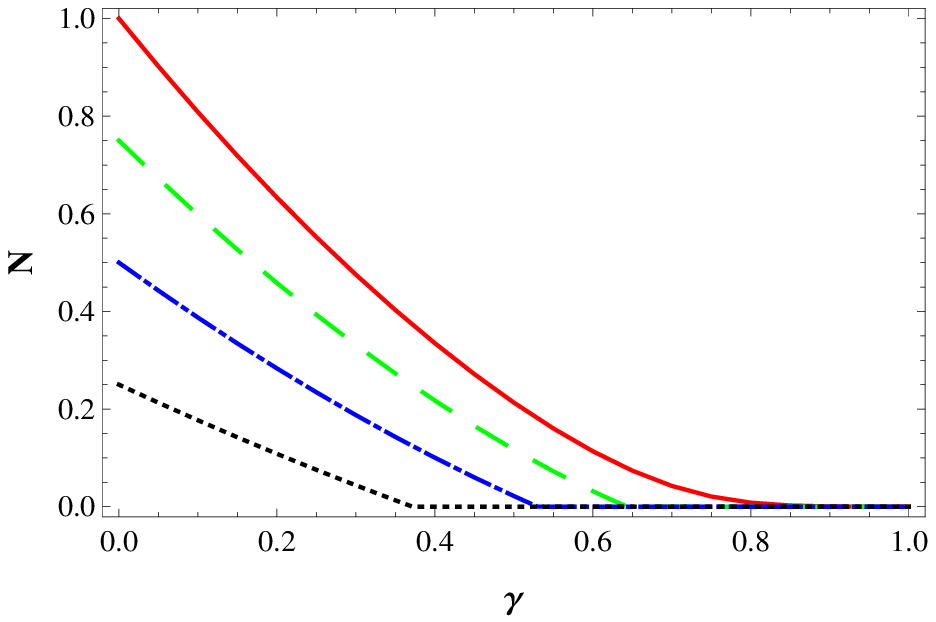}
\caption{Dynamics of entanglement for the system undergoing the
multi-local bit-(trit-) flip noise with parameter $b=0$. The solid,
dashed,  dashed-dotted, and dotted lines correspond to $c=1$,
$c=3/4$, $c=1/2$, and $c=1/4$, respectively.}\label{Fig2}
\end{center}
\end{figure}

\begin{figure}[!h]
\begin{center}
\includegraphics[width=8 cm,angle=0]{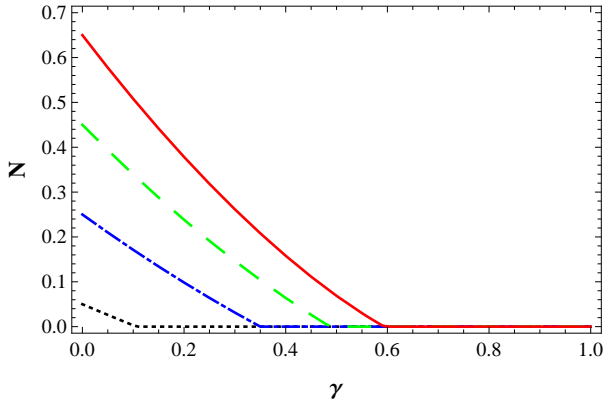}
\caption{Dynamics of entanglement for the system undergoing the
multi-local bit-(trit-) flip noise with $b=1/20$. The solid, dashed,
dashed-dotted, and dotted lines correspond to $c=16/20$, $c=12/20$,
$c=8/20$, and $c=4/20$, respectively.}\label{Fig3}
\end{center}
\end{figure}

As the first example, we consider the initial entangled states shown
in Eq.(\ref{eq.1}) with the parameter $a=0$ which are equivalent to
Werner states \cite{r17}  in a $2 \otimes 2$ systems but with
different noise channels. Their dynamics of entanglement is shown in
Fig.\ref{Fig1}. We note that the time-evolved density matrix for the
initial state with the parameters $a=b=0$ is just the case for the
maximally entangled Bell state $\vert \psi^-\rangle$ \cite{r18},
and the system does not suffer from ESD under a multi-local
bit-(trit-) flip channel.

As a second example, we consider the initial entangled states with
the parameter $b=0$. Their dynamics of entanglement is shown in
Fig.\ref{Fig2}.

Finally, we consider the initial states with $a, b, c >0$. The
dynamics of entanglement is plotted in Fig.\ref{Fig3} for a
particular value of the parameter $b=0.05$.

From the above analysis and numerical results, we conjecture that
ESD in a hybrid qubit-qutrit system under local and multi-local bit-
(trit-) flip channels is a general phenomenon. No ESD takes place
under a qubit-flip channel alone if and only if the qubit-qutrit
system is initially in the states shown in Eq.(\ref{eq.1}) with the
parameter $b=0$. However, ESD always occurs when the system
undergoes a qutrit-flip channel alone.

\bigskip

\subsection{Bit-(Trit-) phase-flip channels}\label{sec3.4} 

The Kraus operators describing the bit-phase flip channel for a
single qubit $A$ are given by
\begin{eqnarray}                                                   \label{eq.23}
E_{1}^{A}&=&\sqrt{1-\frac{\gamma_{A}}{2}}\left(\begin{array}{cc}
1&0\\
0&1\\
\end{array}
\right)\otimes I_{3}, \nonumber \\
E_{2}^{A}&=&\sqrt{\frac{\gamma_{A}}{2}}\left(\begin{array}{cc}
0&-i\\
i&0\\
\end{array}
\right)\otimes I_{3},
\end{eqnarray}
and those for a single qutrit $B$ can be written as
\begin{eqnarray}                                                   \label{eq.24}
F_{1}^{B}&=&I_{2}\otimes\sqrt{1-\frac{2\gamma_{B}}{3}}\left(\begin{array}{ccc}
1&0&0\\
0&1&0\\
0&0&1\\
\end{array}
\right), \nonumber \\
F_{2}^{B}&=&I_{2}\otimes\sqrt{\frac{\gamma_{B}}{6}}\left(\begin{array}{ccc}
0&0&e^{i2\pi/3}\\
1&0&0\\
0&e^{-i2\pi/3}&0\\
\end{array}
\right), \nonumber \\
F_{3}^{B}&=&I_{2}\otimes\sqrt{\frac{\gamma_{B}}{6}}\left(\begin{array}{ccc}
0&0&e^{-i2\pi/3}\\
1&0&0\\
0&e^{i2\pi/3}&0\\
\end{array}
\right),\nonumber\\
F_{4}^{B}&=&I_{2}\otimes\sqrt{\frac{\gamma_{B}}{6}}\left(\begin{array}{ccc}
0&e^{-i2\pi/3}&0\\
0&0&e^{i2\pi/3}\\
1&0&0\\
\end{array}
\right), \nonumber \\
F_{5}^{B}&=&I_{2}\otimes\sqrt{\frac{\gamma_{B}}{6}}\left(\begin{array}{ccc}
0&e^{i2\pi/3}&0\\
0&0&e^{-i2\pi/3}\\
1&0&0\\
\end{array}
\right),
\end{eqnarray}
where $\gamma_{A}=1-e^{-t\Gamma_{A}}$,
$\gamma_{B}=1-e^{-t\Gamma_{B}}$, and
$\gamma_{A},\gamma_{B}\in[0,1]$.

The  elements of the density matrix $\rho$ after the interaction
time $t$ under multi-local bit-(trit-) phase-flip channels are given
by
\begin{widetext}
\begin{center}
\begin{eqnarray}                                                   \label{eq.25}
\rho_{11}(t)&=\rho_{55}(t)=&b+\frac{1}{4}(b-c)\gamma_{A}(\gamma_{B}-1)+(\frac{1}{6}-b)\gamma_{B},\nonumber\\
\rho_{22}(t)&=\rho_{44}(t)=&\frac{1}{12}(6(b+c)-3(b-c)\gamma_{A}(\gamma_{B}-1)+(2-6b-6c)\gamma_{B}),\nonumber\\
\rho_{33}(t)&=\rho_{66}(t)=&\frac{1}{6}(3(1-3b-c)+(9b+3c-2)\gamma_{B}),\nonumber\\
\rho_{15}(t)&=\rho_{51}(t)=&-\frac{1}{12}(b-c)\gamma_{A}(3-2\gamma_{B}),\nonumber\\
\rho_{16}(t)&=\rho_{61}(t)=&\rho_{35}(t)=\rho_{53}(t)=\frac{1}{24}(b-c)(\gamma_{A}-2)\gamma_{B},\nonumber\\
\rho_{24}(t)&=\rho_{42}(t)=&\frac{1}{12}(b-c)(2-\gamma_{A})(3-2\gamma_{B}),\nonumber\\
\rho_{26}(t)&=\rho_{62}(t)=&\rho_{34}(t)=\rho_{43}(t)=\frac{1}{12}(b-c)\gamma_{A}\gamma_{B},
\end{eqnarray}
\end{center}
\end{widetext}
and all the remaining matrix elements are zero.

\textsl{(1) Local bit-(trit-) phase-flip channel only}. We obtain
the same results as the case with a bit-(trit-) flip channel.


\begin{figure}[!h]
\begin{center}
\includegraphics[width=8 cm,angle=0]{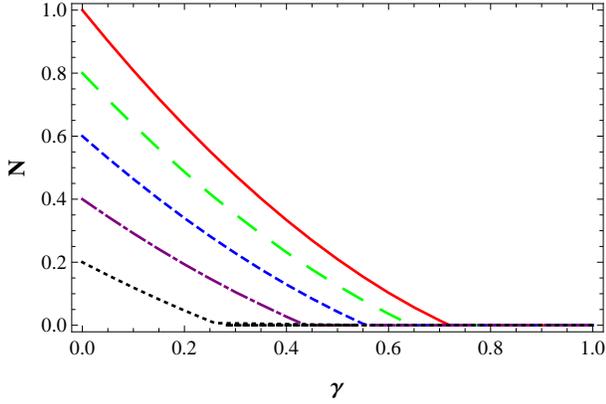}
\caption{Dynamics of entanglement for the system undergoing the
multi-local bit-(trit-) phase-flip noise with  the parameter $a=0$.
The solid, dashed, short-dashed, dashed-dotted, and dotted lines
correspond to $b=0$, $b=1/30$, $b=2/30$, $b=3/30$, and $b=4/30$,
respectively.}\label{Fig4}
\end{center}
\end{figure}

\begin{figure}[!h]
\begin{center}
\includegraphics[width=8 cm,angle=0]{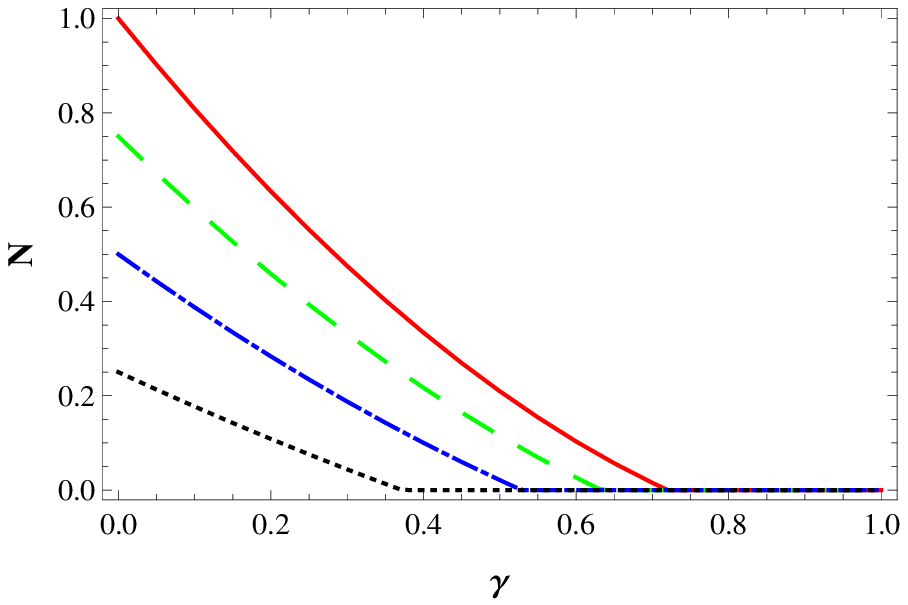}
\caption{Dynamics of entanglement for the system undergoing the
multi-local bit-(trit-) phase-flip noise with the parameter $b=0$.
The solid, dashed, dashed-dotted, and dotted lines correspond to
$c=1$, $c=3/4$, $c=1/2$, and $c=1/4$, respectively.}\label{Fig5}
\end{center}
\end{figure}

\begin{figure}[!h]
\begin{center}
\includegraphics[width=8 cm,angle=0]{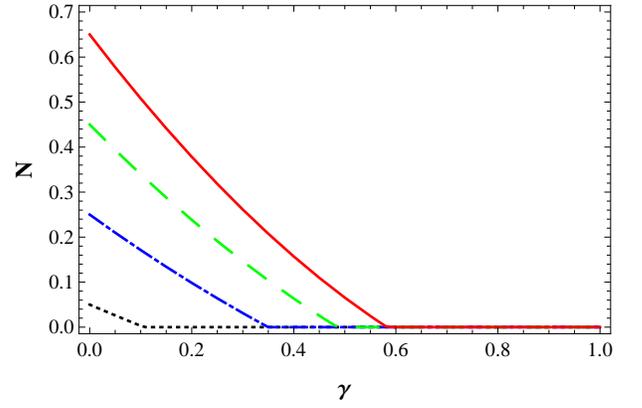}
\caption{Dynamics of entanglement for the system undergoing the
multi-local bit-(tri-t) phase-flip noise with the parameter
$b=1/20$. The solid, dashed, dashed-dotted, and dotted lines
correspond to $c=16/20$, $c=12/20$, $c=8/20$, and $c=4/20$,
respectively. }\label{Fig6}
\end{center}
\end{figure}

\textsl{(2) Multi-local bit-(trit-) phase-flip channel}. For
simplicity, the time dependent parameters are also defined as
$\gamma_{A}=\gamma_{B}=\gamma$. It is  difficult to obtain the
analytical results. The similar work is made as that in the case
with a bit-(trit-) flip channel, and the dynamics of entanglement of
the initial states with the parameter $a=0$ is displayed in
Fig.\ref{Fig4}. Different from a multi-local bit-(trit-) flip
channel, one can see that there exists ESD for the maximally
entangled Bell state (solid line). The dynamics of entanglement of
the initial states shown in Eq.(\ref{eq.1}) with the parameters
$b=0$ and  $a, b, c
> 0$ are displayed in Fig.\ref{Fig5} and Fig.\ref{Fig6}, respectively.

\subsection{Depolarizing channels}\label{sec3.5}

A depolarizing channel represents the process in which the density
matrix is dynamically replaced by the maximally mixed state $I/d$.
Here $I$ is the identity matrix of a single qudit. The set of Kraus
operators that reproduces the effect of the depolarizing channel for
a single qubit $A$  are given by
\begin{eqnarray}                                                   \label{eq.26}
E_{1}^{A} &=& \sqrt{1-\frac{3\gamma_{A}}{4}}I_{6},\nonumber\\
E_{2}^{A} &=& \sqrt{\frac{\gamma_{A}}{4}}\sigma_{1}\otimes I_{3},\nonumber\\
E_{3}^{A} &=& \sqrt{\frac{\gamma_{A}}{4}}\sigma_{2}\otimes
I_{3},\nonumber\\
E_{4}^{A}  &=& \sqrt{\frac{\gamma_{A}}{4}}\sigma_{3}\otimes I_{3},
\end{eqnarray}
where $\sigma_i$ ($i=1,2,3$) are the three Pauli matrices. The Kraus
operators describing a single-qutrit depolarizing noise are given
by\textsuperscript{\cite{19}}
\begin{eqnarray}                                                   \label{eq.27}
F_{1}^{B} &=& I_{2}\otimes\sqrt{1-\frac{8\gamma_{B}}{9}}I_{3},\nonumber \\
F_{2}^{B} &=& I_{2}\otimes \frac{\sqrt{\gamma_{B}}}{3}Y,\nonumber \\
F_{3}^{B} &=& I_{2}\otimes
\frac{\sqrt{\gamma_{B}}}{3}Z,\nonumber \\
F_{4}^{B} &=& I_{2}\otimes \frac{\sqrt{\gamma_{B}}}{3}Y^{2},\nonumber \\
F_{5}^{B} &=& I_{2}\otimes \frac{\sqrt{\gamma_{B}}}{3}YZ,\nonumber \\
F_{6}^{B} &=& I_{2}\otimes \frac{\sqrt{\gamma_{B}}}{3}Y^{2}Z,\nonumber \\
F_{7}^{B} &=& I_{2}\otimes\frac{\sqrt{\gamma_{B}}}{3}YZ^{2},\nonumber \\
F_{8}^{B} &=& I_{2}\otimes \frac{\sqrt{\gamma_{B}}}{3}Y^{2}Z^{2},\nonumber \\
F_{9}^{B} &=& I_{2}\otimes\frac{\sqrt{\gamma_{B}}}{3}Z^{2},
\end{eqnarray}
where
\begin{eqnarray}                                                   \label{eq.28}
Y &=& \left(\begin{array}{ccc}
0&1&0\\
0&0&1\\
1&0&0\\
\end{array}
\right), \nonumber\\ Z &=& \left(\begin{array}{ccc}
1&0&0\\
0&e^{i2\pi/3}&0\\
0&0&e^{-i2\pi/3}\\
\end{array}
\right),
\end{eqnarray}
and $\gamma_{A}=1-e^{-t\Gamma_{A}}$,
$\gamma_{B}=1-e^{-t\Gamma_{B}}$, $\gamma_{A}, \gamma_{B}\in[0,1]$.

The matrix elements of $\rho$ after the interaction time $t$ under a
multi-local depolarizing channel are given by
\begin{widetext}
\begin{center}
\begin{eqnarray}                                                   \label{eq.29}
\rho_{11}(t)&=\rho_{55}(t)=&\frac{1}{12}(12b+3(b-c)(\gamma_B-1)\gamma_{A} + 2(1-6b)\gamma_{B}),\nonumber\\
\rho_{22}(t)&=\rho_{44}(t)=&\frac{1}{12}(6(b+c)-3(b-c)(\gamma_B-1)\gamma_A+ (2-6b-6c)\gamma_B),\nonumber\\
\rho_{33}(t)&=\rho_{66}(t)=&\frac{1}{6}(3(1-3b-c)+(9b+3c-2)\gamma_B),\nonumber\\
\rho_{24}(t)&=\rho_{42}(t)=&\frac{1}{2}(b-c)(1-\gamma_{A})(-\gamma_{B}),
\end{eqnarray}
\end{center}
\end{widetext}
and all the remaining matrix elements are zero.

\textsl{(1) Multi-local depolarizing channel}. The negativity for
the composite system with initial entangled states ($3b<c\leq1-3b$)
is given by
\begin{eqnarray}                                                   \label{eq.30}
N^{mul-loc}(\rho^{AB})=2\max\{0,-\lambda\},
\end{eqnarray}
where
\begin{eqnarray}
\lambda=\frac{9(b-c)\gamma_{A}(\gamma_{B}-1)+2\gamma_{B}(1-9b+3c)+18b-6c}{12}.\nonumber\\
\label{eq.31}
\end{eqnarray}
It is easy to obtain the result that all the states, which are
initially entangled ones, become separable if and only if
$9(b-c)\gamma_{A}(\gamma_{B}-1)+2\gamma_{B}(1-9b+3c)\geq 6(c-3b)$.

\textsl{(2) Qubit depolarizing channel only}. The negativity for all
the states which are initially entangled ($3b<c\leq1-3b$) is given
by
\begin{eqnarray}                                                   \label{eq.32}
N^{bit}(\rho^{AB})=2\max\left\{0,\frac{2c-6b-3\gamma_{A}(c-b)}{4}\right\}.
\end{eqnarray}
The states become separable for all values of $\gamma_{A}\geq
\frac{2c-6b}{3(c-b)}$.

\textsl{(3) Qutrit depolarizing channel only}. The negativity for
all the states which are initially entangled ($3b<c\leq1-3b$) can be
written as
\begin{eqnarray}                                                   \label{eq.33}
N^{trit}(\rho^{AB})=2\max\{0, \frac{3c-9b-(1-9b+3c)\gamma_{B}}{6}\}.
\end{eqnarray}
The states become separable for all values of  $\gamma_{B}\geq
\frac{3c-9b}{1-9b+3c}$.

\begin{widetext}
\begin{center}
\begin{table}[!h]
\tabcolsep 0pt \caption{ESD in $2\otimes3$ systems under certain
noises. } 
\begin{center}
\def\temptablewidth{0.7\textwidth}
{\rule{\temptablewidth}{1pt}}
\begin{tabular*}{\temptablewidth}{@{\extracolsep{\fill}}cccc}
                              & multi-local           & qubit noise only           & qutrit noise only  \\\hline
      dephasing               & ESD with $b\neq 0$    & ESD with $b\neq 0$         & ESD with $b\neq 0$   \\
       phase-flip             & ESD with $b\neq 0$    & ESD with $b\neq 0$         & ESD with $b\neq 0$    \\
       bit-(trit-) flip       & exist ESD             & ESD with $b\neq0$          & ESD  \\
      bit-(trit-) phase-flip  & exist ESD             &ESD with $b\neq0$           & ESD  \\
      depolarizing            & exist ESD             &ESD                         & ESD
       \end{tabular*}
       {\rule{\temptablewidth}{1pt}}
       \end{center}\label{tab1}
       \end{table}
       \end{center}
\end{widetext}

\section{Discussion and summary}\label{sec4}

Putting all the pieces of our results together, one can see that ESD
is a general phenomenon in a qubit-qutrit system undergoing various
independent noise channels and we show the outcomes explicitly in
Table.\ref{tab1}. ``ESD with $b\neq 0$'' denotes  ESD takes places
in a qubit-qutrit system if and only if $b\neq 0$, that is, if b=0,
no ESD occurs.``exist ESD'' denotes ESD may occurs in a qubit-qutrit
system, but not necessary. ``ESD'' denotes ESD always occurs.

Decoherence which is characterized by the decay of the off-diagonal
elements of the density matrix describing the system
\cite{Decoherence},  results from the unwanted interactions of a
quantum system with its environment. According to
Eqs.(\ref{eq.8},\ref{eq.15},\ref{eq.21},\ref{eq.25},\ref{eq.29}),
one can see that the coherence of a qubit-qutrit system can be
destroyed completely when the system undergos the multi-local or
local
 dephasing, phase-flip and depolarizing channels, while
the disappearance of coherence does not occur even though
$\gamma_{A}, \gamma_{B}\mapsto 1$ for multi-local or local the
bit-(trit-)flip and bit-(trit-)phase-flip
 channels.

Using the negativity criterion for quantifying entanglement, it is
shown that the density matrix of a hybrid qubit-qutrit system always
suffers from ESD and decoherence under both local and multi-local
depolarizing channels, which is different from the result that no
ESD in any density matrix of a qubit-qutrit system occurs when only
the qubit is coupled to its depolarizing channel presented in Ref.
\cite{r10}.

The  entanglement dynamics of a specific one-parameter class of
states interacting with a local and multi-lcoal dephasing noise in
hybrid qubit-qutrit systems has been studied by Ann \emph{et al.}
\cite{r9}  in 2008. We have generalized their study to a more
general case, that is, a two-parameter class of states under the
influence of local and multi-local dephasing, phase-flip,
bit-(trit-) flip, bit-(trit-) phase-flip, and depolarizing channels.
With analytical and numerical analysis, we have shown that ESD is a
general phenomenon in a qubit-qutrit system under various noise
channels,
 not only the case with dephasing and
depolarizing ones \cite{r9}.  It can only be avoided in some initial
states undergoing particular noise channels.

In summary, our results show that the  environment, which causes
dephasing, phase-flip, bit-(trit-) flip, bit-(trit-) phase-flip, and
depolarizing, affects the entanglement and the coherence of  a
hybrid qubit-qutrit system in a two-parameter class of entangled
states in very different ways. ESD is a general phenomenon in a
qubit-qutrit system undergoing various independent noise channels
and it can only be avoided in some initial states undergoing
particular noise channels. Moreover, we can divide those noise
channels into two groups. For multi-local and local dephasing,
phase-flip, and depolarizing noise channels, a time scale of
disentanglement is usually shorter than the decay of the
off-diagonal dynamics, and coherence disappears in an infinite-time
limit $t\mapsto\propto$ in which $\gamma_{A}, \gamma_{B}\mapsto 1$.
For multi-local or local bit-(trit-) flip and bit-(trit-) phase-flip
channels, disentanglement may occur in the infinite-time limit, but
the disappearance of coherence does not occur even though
$\gamma_{A}, \gamma_{B}\mapsto 1$.

\section*{ACKNOWLEDGEMENTS}

This work is supported by the National Natural Science Foundation of
China under Grant Nos. 10974020 and 11174039,  NCET-11-0031, and the
Fundamental Research Funds for the Central Universities.

\end{document}